# Common Arc Method for Diffraction Pattern Orientation

Gábor Bortel[*] and Miklós Tegze

*Research Institute for Solid State Physics and Optics of the Hungarian Academy of Sciences, 1525 Budapest P.O.Box 49., Hungary. E-mail: gb@szfki.hu*

**Synopsis** An orientation method based on the intersection arcs of continuous diffraction patterns is described and is shown to perform well on simulated patterns of single particle diffraction experiments performed at x-ray free-electron laser sources.

**Abstract** Very short pulses of x-ray free-electron lasers opened the way to obtain diffraction signal from single particles beyond the radiation dose limit. For 3D structure reconstruction many patterns are recorded in the object's unknown orientation. We describe a method for orientation of continuous diffraction patterns of non-periodic objects, utilizing intensity correlations in the curved intersections of the corresponding Ewald spheres, hence named Common Arc orientation. Present implementation of the algorithm optionally takes into account the Friedel law, handles missing data and is capable to determine the point group of symmetric objects. Its performance is demonstrated on simulated diffraction datasets and verification of the results indicates high orientation accuracy even at low signal levels. The Common Arc method fills a gap in the wide palette of the orientation methods.

## 1. INTRODUCTION

The idea of single particle structure determination by means of diffraction beyond the radiation dose limit using very short pulses of x-ray free-electron lasers has emerged more than a decade ago (Neutze *et al*., 2000; Hajdu, 2000). Although early papers envisaged 3D structure of single biomolecules (Miao, Hodgson & Sayre, 2001; Webster & Hilgenfeld, 2002; Huldt, Szőke & Hajdu, 2003; Miao *et al*., 2004), recent experiments study few hundred nm size particles at most with few nm resolution (Chapman *et al*., 2006; Barty at al., 2008; Mancuso *et al*., 2009; Bogan *et al*., 2010; Loh *et al*., 2010; Chapman *et al*., 2011; Seibert *et al*., 2011). It is true however, that the principle is proven and the progress is continuous. One type of these measurements record large number of continuous diffraction patterns of replicas of the non-periodic object in unknown random orientations and the 3D scattering density is determined through elaborate evaluation processes. The original concept of data processing consists of three separate steps: improving the statistics of low intensity patterns by grouping and averaging, orientation by finding the intersection of the patterns, and finally real space structure reconstruction by iterative phase retrieval. Here we focus on how the unknown random orientation of the individual scattering patterns can be determined.

The method proposed first for orientation originates from the field of cryo-electron microscopy (DeRosier & Klug, 1968; Hart, 1968; Crowther, 1971; van Heel, 1987; Frank, 1996; Penczek, Zhu & Frank, 1996; van Heel *et al*., 2000). There, planar central sections of the 3D Fourier transform of the object are derived from the measured tomographic projections (based on the central section theorem), that are oriented by identifying their straight intersection lines through the origin, the *common lines*. The case of diffraction is different as follows: The measured diffraction patterns define random oriented Ewald spheres in the reciprocal space, so their intersections are circles instead of straight lines. They can be oriented via identifying these *common arcs*; however there are significant differences due to the geometry. Despite knowing the concept of common arcs for a long time (Huldt, Szőke & Hajdu, 2003), the method has never been realized nor analyzed in detail. There is a single case, where the tangent to the common arc – the common line – was utilized to orient diffraction patterns (Shneerson, Ourmazd & Saldin, 2008). This is an approximate solution, which is valid when the curvature of the Ewald sphere is negligible, typically in low resolution, short wavelength measurements. That geometry also necessitates inspection of triplets of patterns, when determining the orientation. Other methods were also proposed for orienting low-intensity diffraction patterns based on generative topographic mapping (Fung *et al*., 2009) and expansion maximization compression (Loh & Elser, 2009) that are able to utilize the similarity of patterns of close orientations and do not require preliminary classification. These methods are shown to be closely related (Moths & Ourmazd, 2010) and, in principle, are able to operate at the lowest intensities close to the limit where the differences due to counting noise or different orientation can be separated (Elser, 2009). A method for obtaining the reciprocal space intensity distribution based on a different principle was also proposed (Saldin *et al*., 2009), and the two approaches were critically compared (Elser, 2010).

In our paper we supply the missing orientation method based on the *common arcs* of continuous diffraction patterns. The method is described in detail, its operation is demonstrated on simulated datasets of two scattering objects, and the results are analyzed carefully. Various questions arisen during experiments, e.g. the question of counting noise, validity of Friedel law, symmetric objects and missing data, are all addressed by the Common Arc orientation method. We obtain rather precise orientations at significantly lower intensities, than it was anticipated (Shneerson, Ourmazd & Saldin, 2008). Also, the complete set of symmetry operations can be determined in case of symmetric scattering objects. The excellent performance is attributed to the exact handling of the curvature of the common arcs and the high redundancy in the complete set of the pairvise determined relative orientations. The elements of the Common Arc orientation method are described in the Methods section and its application is demonstrated in the Results and Discussion section.

## 2. METHODS

### 2.1. Principle of the Common Arc Method

The *Ewald* spheres are 2D spherical slices of the reciprocal space through its origin that represent the achievable region for a single diffraction measurement at constant energy. The Common Arc method determines the unknown orientation of scattering objects by finding the intersection arcs of the corresponding Ewald spheres. The operation of this orientation method can be divided into 2 main steps:

   *i. Determine relative orientation of all pattern pairs by searching for matching common arcs*. The correlation of intensity distribution along intersection arcs as a function of relative orientation is calculated for each pair of diffraction patterns. The best relative orientation is determined by finding the largest correlation. This yields $N(N-1)/2$ unique relative orientations, representing the maximum pairvise obtainable information; $N$ being the number of patterns to orient. However, ultimately we need only $N-1$ relative orientations, as the whole dataset can be arbitrarily oriented. This means $N/2$-fold redundancy, representing the fact that each pattern intersects all other patterns (and vice versa) providing orientation information. This opens up the space for consistency check and averaging of orientations that is performed in the next step.

   *ii. Determine consistent absolute orientation of each pattern by selection and averaging*. The orientation of one arbitrarily selected pattern is fixed and all other patterns' orientation is determined relative to this in several ways utilizing all available relative orientation information. The resulting set of absolute orientations for each pattern is checked for consistency; the reliable ones, which fall close to each other selected and averaged. Then the obtained absolute orientation of each pattern can be used to construct the 3D reciprocal space data for phase retrieval and 3D structure reconstruction.

   In the following sections we discuss the details and the formalism of the Common Arc method.

## 2.2. Relative Orientation

Initially all the scattering patterns are given in the same "laboratory" Cartesian coordinate system, where typically the *xy* "detector plane", is perpendicular to the *z* "beam axis". This data must be located in the reciprocal space by appropriate orientation and projection onto the Ewald sphere. However, without knowing the actual orientation of the sample, we can not properly orient the Ewald sphere. Therefore temporarily all patterns are projected on the same sphere in a standard, laboratory setting. To bring two such projected but not yet oriented patterns ($P_a$ and $P_b$) into a correct relative orientation, where their intersection along the common arc becomes obvious one has to rotate at least one of them. To describe this rotation i.e. relative orientation, we use 3 *Euler* angles, $\Phi$, $\Theta$, $\Psi$ (in the same convention as shown on Figure 1. of Shneerson, Ourmazd & Saldin (2008). The three rotations described by $rot_z(\Phi) \times rot_x(\Theta) \times rot_z(-\Psi)$ are illustrated on Figure 1 and can be explained as follows: One of the patterns in its standard setting (e.g. $P_b$) is rotated about the beam axis by $-\Psi$ azimuthal angle, which brings the tangent to the common arc through the origin (i.e. the common line) of this pattern to the *x* axis. Then this pattern is tilted about the *x* axis by $\Theta$ angle. We call this $\Theta$ angle the hinge angle, as this defines the "inclination" of the two intersecting Ewald spheres and ultimately determines the curvature of the common arc. Finally, the tilted pattern is rotated again about the beam axis by $\Phi$ azimuthal angle to bring the common arc of $P_b$ into an exact overlap with that of $P_a$ still being in its standard setting. In an other view, the inverse of this last rotation could be equivalently applied to the $P_a$ pattern. Although the Euler angles are not always the best choice for orientation/rotation parameterization (vs. e.g. quaternion representation), in this case they exactly correspond to the beam axial rotation of the two patterns and their actual inclination. Therefore we find them the most suitable to the problem.

After performing the above rotations we have set the two scattering patterns into correct relative orientation as it is illustrated on Figure 2. The two spheres are the intersecting Ewald spheres; the areas with polar grid illustrate the measured regions of the two scattering patterns. Their intersection, the common arc (CA) is plotted in red and the tangent to the common arc through the origin (O) is the approximating common line (CL) is shown in green. It can be seen how limited is the scattering range, where the common line well approximates the common arc.

## 2.3. Common Arc

In order to be able to correlate intensities along the common arc we have to determine its equation. It follows from the parameterization of relative orientations by the Euler angles, that the $\Phi$ and $\Psi$ rotations define only the azimuthal position of the common arc in the $P_a$ and $P_b$ patterns. The curvature of the common arc is defined solely by the $\Theta$ hinge angle. This geometry suggests polar gridding of the patterns on the Ewald sphere and polar parameterization for the equation of the common arc. To avoid confusion with the Euler angles or the Bragg angle, let's $\mu$ denote the *polar* and $\nu$ the *azimuthal* angular coordinate. $\mu$ runs from the minimal to the maximal scattering angle, its zero corresponds to the forward scattering direction and optimally $\nu$ covers a whole circle. After a simple and here not detailed geometric calculation based on right-angled triangles on two perpendicular projections of Figure 2, the following base equation is derived for the common arc in the polar detector coordinates:

$$\sin \nu = \frac{1 - \cos \mu}{\sin \mu} \tan \frac{\Theta}{2} \qquad (1)$$

It is plotted in red on Figure 3 for several $\Theta$ angles to illustrate how the common arc bends starting from a line, becomes a closed circle and finally shrinks to a point as $\Theta$ runs from 0 to $\pi$. The formula has the following properties: The origin, $\mu = 0$, $\nu = 0$ is always an asymptotic solution to this equation; this is a fix point to all common arcs. Apart from the $\Theta = 0$ and $\Theta = \pi$ singular cases, there are two symmetric branches starting from the origin: $\nu$ and $\pi - \nu$ representing the two halves of the common arc. Depending on $\Theta$, these two branches meet again at µ = π – Θ polar and ν = π / 2 azimuthal angle, and the common arc becomes a full common circle. If Θ = 0 (the two Ewald spheres exactly overlap), or if Θ = π (the two Ewald spheres face each other), the common arc becomes degenerate (extends to the whole sphere or shrinks to a point). These singular cases are not described by the above equation. It is more important however, that this equation defines the arcs in the two investigated patterns along which we will have to find similar intensity distributions. It is done with an exhaustive 3D search of the Θ, Φ and Ψ angles that yield the maximum correlation. Although the above base equation depends on the Θ hinge angle only, it is assumed that the Φ and Ψ azimuthal rotations already has been applied to the two intersecting patterns. These rotations by definition simply just shift the ν azimuthal coordinates, so the polar coordinates of the common arc will be (µ, Φ – ν) and (µ, Ψ + ν) in the $P_a$ and $P_b$ patterns respectively. Note the opposite signs of ν due to opposite curvature of arcs in the two patterns.

To easily obtain the data points along the common arc, it is expedient to resample the measured patterns to a polar grid by some kind of interpolation and/or averaging during preprocessing. (At least one transformation of the data is likely to be unavoidable for any data evaluation, as the typical Cartesian 2D detector pixel arrangements result a hard-to-use distorted grid on the Ewald sphere, which is inconsistent with our ultimately desired 3D Cartesian grid in the reciprocal space required by iterative reconstruction methods.) This way the advantageously chosen parameterization and polar gridding of the patterns together make possible a fast implementation of the 3D maximum search procedure.

## 2.4. Friedel Law

If the scattering process involves no phase shift, it can be described by real scattering factors and the *Friedel* law applies: The intensities in the reciprocal space possess inherent inversion symmetry, since the corresponding scattering amplitudes are complex conjugates. For light elements, the Friedel law is applicable in most of the x-ray scattering cases, if we avoid the anomalous scattering near absorption edges. However, it becomes definitely invalid if the energy is lowered to the soft-x-ray region, for-example to utilize transparency in the water window.

If the conditions of the measurement make the Friedel law applicable, that will advantageously affect our common arc orientation method, as follows. Due to the inversion symmetry, any two opposite points in the reciprocal space will have the same intensity. Consequently, any measured scattering pattern determines the intensity distribution in the reciprocal space on two Ewald spheres: on one sphere directly and on the opposite sphere through the inverted pattern. We can exploit this when searching for common arc of $P_a$ and $P_b$ patterns by using their opposite patterns $\overline{P}_a$ and $\overline{P}_b$, and simultaneously compare intensities along the intersection arc of $P_a$ and $\overline{P}_b$, for example (other combinations would not yield independent information). This effectively doubles the average length of the correlated arcs, making the correlation factor more reliable, less sensitive to noise.

The "Friedel" common arcs are shown on both Figure 2 and Figure 3 in blue. Their equation can also be calculated using the above formula (1), except the $\tan(\Theta/2)$ must be replaced by $-\cot(\Theta/2)$ due to the inversion of one of the spheres. For the same reason, the Friedel common arcs are flipped. Figure 3 illustrates the complementary role of the two sets of arcs.

The option to include the Friedel common arcs into the relative orientation search and forcing the inversion symmetry when the 3D reciprocal space intensity is constructed from the oriented individual scattering patterns depends only on the physical applicability of the Friedel law and in no other way influences the process of orientation.

## 2.5. Correlation

When searching for the Φ, Θ, Ψ Euler angles determining the best matching relative orientation of two patterns along their common arcs, a weighted *Pearson* correlation factor is calculated as a measure of the similarity. This takes the following form:

$$\chi = \frac{\sum_{l=1}^{L} w_l (a_l - \bar{a})(b_l - \bar{b})}{\sqrt{\sum_{l=1}^{L} w_l (a_l - \bar{a})^2} \sqrt{\sum_{l=1}^{L} w_l (b_l - \bar{b})^2}}, \quad \bar{a} = \frac{\sum_{l=1}^{L} w_l a_l}{\sum_{l=1}^{L} w_l}, \quad \bar{b} = \frac{\sum_{l=1}^{L} w_l b_l}{\sum_{l=1}^{L} w_l} \quad (2)$$

Here all sums run through the $L$ pixels along the common arcs (either including the Friedel common arcs or not), $a_l$ and $b_l$ denote the interpolated intensities from the two compared patterns and $w_l$ implements an optional weighting. These are obtained as follows. The scattering patterns already given on the polar grid described above are prenormalized with the average scattering angle dependence of all patterns. This prevents the domination of often orders of magnitude higher intensities at low scattering angles. At the same time, their higher reliability due to counting statistics can be taken into account by properly chosen $w_l$ weights. These weights can also express the higher sensitivity of the outer pixels to the orientation. The exact weighting scheme is a parameter of the method, which typically can be a function of intensity, statistics or even the scattering angle. However, regardless of the scale of the correlated values and the applied weighting, the correlation factor falls in the [−1,+1] range, −1 representing anticorrelation, 0 no correlation and +1 the perfect correlation. This puts the $\chi(\Theta, \Phi, \Psi)$ correlation maps on an absolute scale, when searching for the maximal $\chi$ value on a discrete grid of the Φ, Θ, Ψ Euler angles. This is useful in the judgment of the best matching common arc of each pattern pair.

## 2.6. Absolute Orientation

Once the relative orientation of all pattern pairs described by the rotation matrix $R_{ij}$ is found, an absolute orientation matrix, $O_i$ has to be determined for each pattern. For one selected pattern, e.g. $O_1$ it can be arbitrary, e.g. the eye matrix, as the orientation of the whole reciprocal space is irrelevant. Then, the orientation of all other patterns is obtained by: $O_i^{(1)} = R_{i1} \times O_1$. In principle this step solves the orientation problem, as all orientations are determined. However, one should go further to exploit the high degree of redundancy in the complete set of relative orientations. This is achieved by successive application of two relative rotations: $O_i^{(j)} = R_{ij} \times R_{j1} \times O_1$, to utilize all intermediate rotations. The above equations yield $N-1$ candidate orientations for each pattern. In an ideal case these would be identical, but due to the sampled, interpolated and noisy data and the discrete grid used to search the relative orientations, they scatter, or even can be completely erroneous, i.e. identify a false common arc. Therefore, a "*selection and averaging*" procedure is performed on the $N-1$ candidate orientations to improve the reliability and precision of the final absolute orientation assigned to the given pattern. This selection and averaging is illustrated on the left panel of Figure 4, and is realized as follows. *i*. For each pattern, the distances of candidate orientations is calculated as $\delta_{jj'} = \arccos((\text{trace}(O_i^{(j)} \times O_i^{(j')}) - 1)/2)$. The misorientation angle given by this formula is a natural metric in the rotational group (Morawiec, 2004). *ii*. For each candidate, $O_i^{(j)}$ the number of other candidates, $O_i^{(j')}$ falling within a distance range given by a tolerance angle, $\delta_{\text{lim}}$ is determined. *iii*. The most populated neighborhood is selected; in case of a tie, the smaller average misorientation is preferred. *iv*. The orientations within this group of candidate orientations is averaged and assigned to the given pattern as its most probable absolute orientation, $O_i$. The selection within a range eliminates outliers, and the averaging improves the precision. The quality of the result is ensured by the proper choice of the tolerance angle parameter, which should allow averaging of naturally scattered orientations but should exclude erroneous ones.

## 2.7. Symmetric Objects

The structure of single particles one wishes to determine may be symmetric. One can think of simple geometric body-shaped nanoparticles (Chapman *et al.*, 2011), icosahedral viruses (Seibert *et al.*, 2011) or the quaternary structure of multi-chain proteins. The real space symmetry of the scattering density given by the point group leads to at least the same symmetry of the Fourier intensities in the reciprocal space. An inversion is possibly added (if not present before), due to the Friedel law, as discussed above. From the point of view of orientation problem symmetry means, that a given pattern's orientation can not be unambiguously determined: The same measured pattern can be located in several *equivalent settings* in the reciprocal space, related by unknown symmetry operations. Note, that we use the word setting, instead of orientation, as they may be related by other operations than a simple rotation (see below).

The Common Arc orientation method was extended not just to handle scattering patterns of such symmetric objects, but also take advantage of it. With this addition it is possible to identify the symmetry elements and also their orientation (the direction of rotation axes and normal vectors of mirror planes) in the reciprocal space. The idea is the following: Equivalent settings of the same pattern will yield several equally high peaks in the $\chi(\Theta,\Phi,\Psi)$ correlation map, when searching for the matching common arcs as a function of the relative orientation of two patterns, $P_a$ and $P_b$. Each of these peaks indicates a highly correlating common arc and its position in the map provides the appropriate relative rotation. As a complication, equivalent settings can be related by other operations than a proper rotation. These are the inversion, a reflection or an improper rotation. However, these all can be separated to an inversion and a proper rotation. The latter component can be found by doubling the correlation map, i.e. calculating $\overline{\chi}(\Theta,\Phi,\Psi)$ using one of the patterns inverted, $\overline{P_b}$, for example. If reflective symmetry is present, this map contains the same number of peaks with similar quality as the first map; otherwise it contains no peaks at all. The total number of peaks found on the two maps ($\chi$ and $\overline{\chi}$) should be equal the number of equivalent settings, i.e. the *order* or *multiplicity*, $M$ of the point group symmetry of the reciprocal space. By defining appropriate criteria on the local maxima on the maps this number can be determined automatically. However, in the present implementation of the algorithm, the expected multiplicity is given as an input parameter and that number of peaks is identified on the correlation maps.

Once the $M$ relative operations (rotations, or possibly rotations combined with inversion), $^sR_{ij}$ are determined for all pattern pairs from the correlation maps we derive the absolute settings of the patterns based on similar equations that we used in the absence of symmetry: $^sO_i^{(1)} = {}^sR_{i1} \times O_1$ and $^{s,s'}O_i^{(j)} = {}^sR_{ij} \times {}^{s'}R_{j1} \times O_1$. The $s$ superscript refers to one of the $M$ symmetry equivalents. The only difference is that some of these equivalent settings may involve inversion and also in that we will obtain a total $M + M^2 \times (N-2)$ of them. This set of candidates contains all the $M$ equivalent absolute settings mixed together, with a redundancy on the order of approximately $M \times N$-fold. Therefore, a kind of clustering must sort them out and the redundancy is exploited by averaging. We can apply again the above described selection and averaging procedure $M$ times repeatedly, and remove the already averaged elements between the runs. The procedure is illustrated on the right panel of Figure 4. This approach resembles the *quality threshold clustering* algorithm (Heyer, Kruglyak & Yooseph, 1999) in that sense it limits the extent of clusters, but here the number of clusters is also limited. Consequently, some elements will not belong to any of the clusters; these are treated as outliers. By obtaining $M$ averaged absolute settings for each pattern the orientation problem of symmetric object is solved, but we can go further and determine the symmetry operations themselves.

Any operation that transforms one setting of a pattern to an equivalent one is a *symmetry operation*. These can be written with the help of the above determined absolute orientations as $^sO_i \times ({}^{s'}O_i)^{-1}$. This set contains $M^2 \times N$ symmetry operations derived from the equivalent settings of all patterns. However, many of them should be close to others, as there are only $M$ common symmetry operations for all patterns, representing again an $M \times N$-fold redundancy. Their mixture is clustered into $M$ distinct, quality-assured groups again by the repeatedly applied selection and averaging procedure to provide the $M$ averaged symmetry operations, i.e. the elements of the point group of the reciprocal space. The special directions of these operations (axes of rotations and normal of reflections) reveal the high-symmetry directions of the oriented patterns, which could be very useful during a visual inspection. Also, and perhaps more importantly, the relation of the determined symmetry operations can be analyzed to verify whether they form a valid point group. If not, possibly one can get a clue to cancel or include some symmetry elements. However, this goes beyond the scope of this paper.

One may argue against the implicit treatment of the Friedel inversion symmetry as described earlier, saying that inversion is just a symmetry element and could be treated the same way as any other reflective symmetry element described in this section. However, there is a significant difference, which makes inversion unique: It is the only symmetry element that can be applied in lack of orientation information. It requires only fixing the origin, in contrast to other type of point group symmetries that involve some unknown orientation of the symmetry element (mirror plane or rotation axis). Therefore, the latter symmetry elements can only be treated as described here. While the inversion could be treated both ways, it is preferable to handle implicitly during the correlation mapping as it improves the reliability of the determined relative orientations.

### 2.8. Missing Data

An ideal single particle diffraction pattern would cover the whole annulus between the coaxial cones defined by the beamstop and the crystallographic resolution. However, the patterns obtained experimentally often have unmeasured, or not useable regions even within this area. These missing regions are typically due to gaps between elements of a multidetector system, due to saturation at low angles, bad pixels or any other experimental deficiency. In general, an arbitrary mask on the data must be handled.

Missing data can be treated when calculating the above defined correlation factor, simply by using only the pixels along the common arc present in both patterns ($l \in P_a \cap P_b$). In principle that is all has to be done, however, this makes the calculation more demanding as the number of common pixels and the normalization varies with the relative orientation. Yet to be able to prenormalize the patterns, as an *approximation*, we have slightly modified the formula for the correlation written as the ratio of the covariance and the product of standard deviations:

$$\chi \approx \frac{\sum_{l \in P_a \cap P_b} w_l (a_l - \bar{a})(b_l - \bar{b}) \Big/ \sum_{l \in P_a \cap P_b} w_l}{\sqrt{\sum_{l \in P_a} w_l (a_l - \bar{a})^2 \Big/ \sum_{l \in P_a} w_l} \sqrt{\sum_{l \in P_b} w_l (b_l - \bar{b})^2 \Big/ \sum_{l \in P_b} w_l}}, \quad \bar{a} = \frac{\sum_{l \in P_a} w_l a_l}{\sum_{l \in P_a} w_l}, \quad \bar{b} = \frac{\sum_{l \in P_b} w_l b_l}{\sum_{l \in P_b} w_l}$$

(3)

Here, the weighted covariance is calculated for the pixels of the common arc present in both patterns, but the weighted mean and standard deviations are calculated for the pixels present in the given pattern. This approximation assumes that the weighted mean and standard deviation is not affected very much by the elements present only in one, but not in the other pattern. It is acceptable, if a moderate fraction of data points is missing. The formula returns the exact value if there is no missing data.

## 3. RESULTS AND DISCUSSION

### 3.1. Preparation of Synthetic Data

Beyond description of the Common Arc orientation method, the secondary object of our work is to demonstrate its capabilities in finding the correct orientation of single particle diffraction patterns. To be able to *validate* the results the operation is demonstrated on synthetic, but as realistic as possible data created under total control of parameters. This is achieved in three separate groups of tasks: *i.*, Calculate numerous diffraction patterns of selected test objects in random orientation. Make them more realistic by including counting noise and introducing missing regions in the patterns. *ii.*, Transform the patterns from the Cartesian detector grid to a polar grid suitable for the orientation algorithm. Perform the Common Arc method: Map the correlations with relative orientation, determine the absolute orientation of each pattern and the symmetry operations, if applicable. Also combine the oriented 2D patterns to the 3D reciprocal space distribution ready for structure reconstruction. *iii.*, Verify the obtained orientations against the ones used at the preparation phase and analyze the errors that characterize the method.

To demonstrate and test all aspects of the Common Arc method we have chosen *two scattering objects*. One of them was inspired by recent experiments performed at the already operational x-ray free-electron lasers (Seibert *et al.*, 2011). It is a model of a large virus with internal structure and pseudo symmetry, investigated with non-atomic resolution using longer wavelength radiation and the Friedel law is assumed to be invalid. The other one is an example of objects targeted by the ultimate single particle diffraction experiments planned at future x-ray free-electron laser sources (Miao, Hodgson & Sayre, 2001). It is a symmetric protein structure investigated with atomic resolution using shorter wavelength radiation and the Friedel law is assumed to be valid. The details of the pattern-generating procedure are described below and the important parameters of these two objects and the simulated experiment are listed in Table 1.

*i. Virus Model with Internal Structure and Pseudo Symmetry.* The overall shape of the model object was created with the help of Gielis curves (Gielis, 2003) extended to two dimensions, also called spherical product of two superformulas. These functions define simple 2D surfaces with the help of few control parameters resembling the shape of various bodies found in nature. This boundary was smeared out with the help of a Fermi-Dirac function yielding the primary 3D scattering density. With the linear combination of several such primary densities it is possible to create some lower or higher density regions within the particle or even features such as a shell with different density. Then a 20 % random fluctuation of this combined scattering density was introduced to imitate a fine local structure with relatively low contrast. Finally, an imaginary part was added to the density in order to yield scattering patterns not obeying the Friedel law. Its amplitude was set to 10 % of the real density according to the typical $f_2/f_1$ ratio of the atomic scattering factor corrections of C, N and O elements at the given radiation energy. The absolute value of the complex scattering density is illustrated on Figure 5.

*ii. Symmetric Protein Molecule at Atomic Resolution.* Our choice fell on the Ribulose 1,5-bisphosphate carboxylase/oxygenase protein assembly, RuBisCo (Wildman, 2002; Portis & Parry, 2007) partially because of its abundance in the related literature (Miao, Hodgson & Sayre, 2001) including our study on grouping of its low-intensity patterns (Bortel & Faigel, 2007; Bortel, Faigel & Tegze, 2009) and partially because of its highly symmetric tertiary structure. Structural data was taken from the Protein Data Bank (PDB) ID: 1EJ7 (Duff, Andrews & Curmi, 2000), and the biologically active unit of 8 long and short chains was generated using the 422 point group symmetries. The weight of the approximately spherical assembly is about 0.54 MDa. According to the high energy of the radiation, the Friedel law was assumed to be valid.

Calculation of the scattering amplitudes was somewhat different for the two objects. For the 3D electron density model the scattering amplitude in the reciprocal space was calculated by 3D discrete Fourier transform. The simulated scattering patterns on the planar Cartesian detector were obtained by interpolation to the corresponding 2D projected grid of the random-oriented Ewald spheres. For the atomistic protein model the scattering amplitudes at each scattering vector of the 2D pattern were calculated directly by summing up the spherical wave contributions of each atom, taken into account the atom's form factor and its phase due to its position. In case of both objects the scattering amplitudes were squared and *scaled* to photon counts that enabled us to include counting statistics into the simulated scattering patterns. For this we assumed a realistic 0.4 e/Å$^3$ average electron density in the case of virus model object, experimentally achievable photon fluence (number of incoming photons per focal spot area) and detector pixel size corresponding to the Shannon-Nyquist sampling. The solid angle of this pixel was taken as $(\lambda/2D)^2$, $\lambda$ being the wavelength and $D$ being the object size (Huldt, Szőke & Hajdu, 2003; Shneerson, Ourmazd & Saldin, 2008).

As the last step of synthetic data creation, a random 0.1 % of the pixels and the ones having counts larger than 1 % of the forward scattering value were masked out, representing some bad and saturated pixels, respectively. One of the synthetic scattering patterns of the RuBisCo enzyme is shown on Figure 6. Note, that in spite of the valid Friedel law, the individual scattering patterns are not centrosymmetric, due to the curvature of the Ewald sphere. Nevertheless, an approximate symmetry is observable in the central region, where the tangent plane well approximates the sphere. A total of 100 patterns were generated for both objects and passed to the orientation procedure, keeping the orientation information for verification purposes.

### 3.2. Orientation by the Common Arc method

The scattering patterns created on a Cartesian detector grid are first transformed onto a polar grid suitable for the common arc search procedures. This extra step was intentionally included as a preparation to accept *real data*. The different geometry of the two grids, specifically the mismatch of the point densities necessitated a combined interpolation and averaging procedure. To utilize most of the information, interpolation is better choice, where the targeted polar grid is denser than the original Cartesian grid, and averaging is preferable in the opposite case. The step size of the polar grid was chosen to 0.5º in the polar and 1º in the azimuthal direction. Consequently, the steps of the 3 Euler angles during the orientation search were also 1º. In the correlation factor calculations all pixel values were normalized with their corresponding solid angle, i.e. the intensity was used rather than the pixel size dependent counts. Furthermore, all patterns were prenormalized by the average scattering angle dependence to prevent domination of the low scattering angle, high intensity regions. The weights used in the correlation factor were chosen proportional to the scattering angle representing a higher sensitivity of the outer pixels to the orientation. The validity of Friedel law and an 8-fold symmetry multiplicity was assumed in the case of RuBisCo. The Common Arc orientation of 100 patterns with the above parameters takes several hours on a typical desktop computer. The most time consuming task is to calculate the correlation map of the relative orientations. It could be speeded up by parallelization.

The success of the Common Arc orientation process can be assessed from several *indicators* already available during its progress. The largest correlations found, indicate the match of the intensities along the common arcs. At the lowest counting statistics of the patterns the values above 0.5 may already indicate a correctly identified intersection, but at high pattern intensities the correlation can reach even 0.99. This indicates a minor contribution to the errors from the discrete gridding; the non-perfect correlation can be attributed mostly to the shot noise. It is also worth to analyze the amplitude, shape and width of the peaks surrounding the local maxima found in the $\chi(\Theta,\Phi,\Psi)$ maps. This helps to optimize the search procedure, namely determine the necessary step size of the maps. Figure 7 shows a perfectly matching common arc of 2 patterns of the virus model object found by the algorithm. The corresponding correlation factor of the intensities along the arc is 0.95.

When determining the absolute orientation by the selection and averaging procedure, the *fraction* of selected and averaged orientations serves as another useful indicator on the quality of the result. Its higher value expresses the consistency of the candidate orientations and the reliability of the average. A value of 100 % would indicate that all common arcs have been consistently found and contribute to the average. Nevertheless, our experience shows that a value as low as 10 % is already enough to correctly orient all patterns with low error (see the verification, below). The relevant parameter, the averaging range angle, $\delta_{\text{lim}}$ was chosen as

3°, few times the grid step size, which was confirmed by the typical peak widths found on the orientation map and present in the misorientation angle distributions. This shows the importance and effectiveness of this step in filtering the consistent orientations.

In case of symmetric objects, such as the RuBisCo, the symmetry operations determined by the Common Arc method also give the opportunity of a self-consistency check. Figure 8 illustrates how the symmetry operations are obtained: Two equivalent settings of one selected example pattern defined by multiple peaks of the $\chi(\Theta, \Phi, \Psi)$ maps reveal the underlying symmetry operation, in this case a 180° rotation indicated by the red arrow. If these operations are found to be common for the equivalent setting of all patterns, then ultimately they define the corresponding 2-fold symmetry axis with high reliability via selection and averaging. Other symmetry operations can be derived similarly, a total of 8, the expected multiplicity. The axes of these operations are also shown on Figure 8. Their folds were determined from the corresponding rotation angles that happened to be 0°, 90° and 180°, which is already a good sign. Even more, the symmetry axes make 45°, 90° or 135° angles. It can be verified that these symmetry operations do form the complete 422 point group. In general, finding symmetry elements that define a valid point group is a strong indication of the correctly identified symmetry group, merely from its multiplicity. As a control trial, we attempted to perform the Common Arc orientation by assuming a multiplicity of 7 or 9. The result was either lack or a surplus of proposed symmetry operations. However, the required completeness of the point group easily revealed the correct symmetry. The high accuracy of the symmetry operations determined from the equivalent settings of all patterns – less than 0.01° error in the rotation angles and the direction of axes – is attributed to the high degree of redundancy.

### 3.3. Verification of the Results

The above observations characterize the Common Arc orientation method in itself and can also be examined in case of real data. Now, taking advantage of synthetic data, we can judge the results based on the original orientation information kept from the pattern-preparation phase. This allows us to quantitatively verify the results and specify the operation limits and the effectiveness of the method.

Figure 9 shows the distribution of *misorientation* angles of the original and the oriented patterns for both scattering objects. The 0.4° and 0.1° mean values convert to 1.5 and 0.4 pixel positional error at the perimeter (the highest resolution ring) of the individual scattering patterns. This means that in both cases all patterns have been oriented by the Common Arc method precisely enough for structure determination to high resolution. It has to be noted that more patterns or better statistics would further decrease the orientation errors.

This naturally raises the question of intensity that is required for successful orientation. The incoming x-ray pulse fluences listed in Table 1 represent safe operation limits. Lower values already caused significant orientation errors for a few percent of the patterns, however, large fraction of the patterns was still correctly oriented. In case of virus model object the stated fluence is well above the parameters of the already operational FLASH and LCLS xfel sources (FLASH; LCLS) it is only 3% of $10^{12}$ photons focused to $10\times10$ μm$^2$. However, assuming $10^{12}$ photons focused to $100\times100$ nm$^2$ from the European XFEL source (XFEL), successful orientation of the RuBisCo patterns requires preliminary grouping and averaging, the so called classification of patterns (Huldt, Szőke & Hajdu, 2003; Bortel & Faigel, 2007; Bortel, Faigel & Tegze, 2009), with an average 30 patterns/group. This reduces the effective minimal fluence required by the combined classification and orientation process.

For easy comparison to the work of Shneerson, Ourmazd & Saldin (2008) on the orientation by the common lines, the mean photon count in a Shannon-Nyquist pixel at the highest resolution ring is also given in table 1. The sufficient 0.7 and 0.2 counts for our 2 scattering objects indicate that the Common Arc method is able to orient at significantly *lower intensities* with *higher precision* than it was anticipated from the performance of the approximate method. The cited work states 10 counts per pixel and 4° average misorientation error and draws a rather pessimistic conclusion on the feasibility of such experiments. The origin of the improvement is the precise handling of the curvature of the common arcs and exploiting the high degree of redundancy of all relative orientations. This signal level also gets close to that of other algorithms (Loh & Elser, 2009; Fung *et al*., 2009).

The iterative phase retrieval algorithms – the last step in determining the actual density of the scattering object – typically require the reciprocal space intensities on a 3D Cartesian grid, suitable for Fast Fourier Transform cycles. This data was prepared by binning all original 2D patterns (not the ones transformed to the polar grid) to the voxels of the 3D grid utilizing the orientation information provided by the Common Arc method. Contributions to each voxel were then averaged. 3 main slices of this volume data through the origin of the reciprocal space are shown on Figure 10. The outer regions with no pattern-contribution yield a fuzzy surface of the roughly spherical volume that is defined by the original diffraction patterns. Intensity of the voxels further in, with contributions from many patterns possesses very high reliability. Although this volume contains all pixels of all patterns, it obviously has to be trimmed to make it useable by the reconstruction process. The voxel-voxel correlation of this data and the original intensity distribution in the reciprocal space calculated directly from the scattering density was determined as another quantitative measure of the accuracy. The 0.98 correlation factor indicates that the Common Arc orientation, including the necessary data transformation from the 2D detectors to the 3D volume grid introduces minimal errors to the whole structure determination process.

**4. CONCLUSION**

The Common Arc method for orientation of continuous diffraction patterns of non-periodic single particles has been described in detail. The method relies on the curved intersection of Ewald spheres in finding the relative orientation of all pairs of diffraction patterns, and then determines a consistent orientation for each of them via selection and averaging. The approach is significantly new compared to recent application of the approximate method of common lines (Shneerson, Ourmazd & Saldin, 2008) adapted from the field of cryo-electron microscopy. Taken simulated diffraction patterns of various scattering objects, the Common Arc orientation method is shown to be able to operate at significantly lower intensity levels (less than a count per pixel) and yield orientations with higher precision (within few tenths of a degree) than it was anticipated earlier. Also it was extended to handle missing regions in the diffraction patterns and to take advantage of the Friedel symmetry, if exists. In case of symmetric scattering objects the symmetry operations are automatically obtained from the equivalent settings of the patterns and ultimately the complete point group symmetry can be determined. It is a non-iterative method providing deterministic results, with various indicators on the reliability and consistency of the orientations, therefore it can also aid iterative orientation algorithms with a biased state in emerging from their random initial states. It is also shown that this orientation process introduces no artefacts into the 3D reciprocal space intensity distribution.

**Acknowledgements**   This work was supported by the Hungarian Research Fund, OTKA grant numbers K81348 and K67866. G. B. acknowledges support of the Bolyai János Scholarship of the Hungarian Academy of Sciences.

**Table 1** Parameters of the two scattering objects used in preparation of the simulated single particle diffraction experiment.

| | virus model | protein molecule |
|---|---|---|
| size of the object | ~ 5000 Å | ~ 150 Å |
| symmetry of object | pseudo 5-fold axis | exact 422 point group |
| wavelength / energy of radiation | 50 Å / 248 eV | 1.24 Å / 10 keV |
| Friedel law | invalid | valid |
| pulse fluence | $0.03 \times 10^{12}/(10\ \mu m)^2$ | $30 \times 10^{12}/(100\ nm)^2$ |
| crystallographic resolution | 100 Å | 2.4 Å |
| maximal scattering angle | 29º | 30º |
| counts in outer Shannon-Nyquist pixels | 0.7 | 0.2 |
| fraction of missing pixels | 0.22 % | 0.15 % |
| number of patterns | 100 | 100 |

$rot_z(-\Psi)$   $rot_x(\Theta)$   $rot_z(\Phi)$

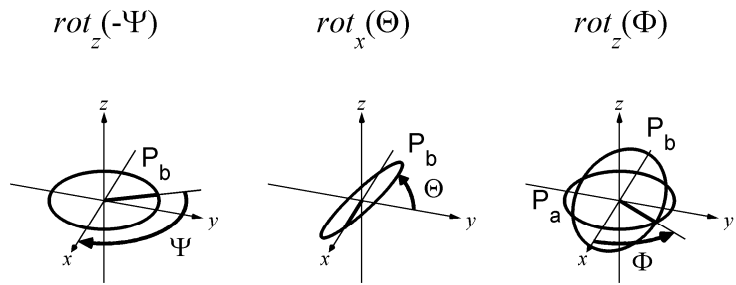

**Figure 1** Illustration of the 3 Euler rotations describing the relative orientation of two patterns, $P_a$ and $P_b$. Circles indicate either the tangent planes to the Ewald spheres at the origin of the reciprocal space or equally the detector planes in the real space. For more explanation see text.

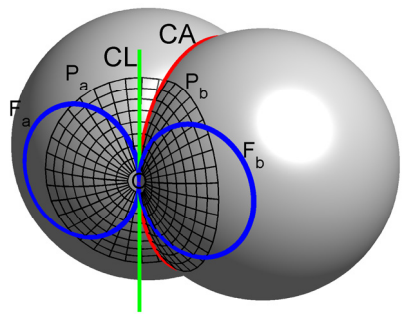

**Figure 2** Intersection of two Ewald spheres showing the geometry of two diffraction patterns ($P_a$, $P_b$), their common arc in red (*CA*), it's tangent at the origin, the approximating common line in green

(*CL*) and the common arc of a pattern with the other pattern's Friedel pair in blue ($F_a$, $F_b$). The Friedel pairs of the Ewald spherers, i.e. their inversion across the origin (*O*) are not shown for clarity.

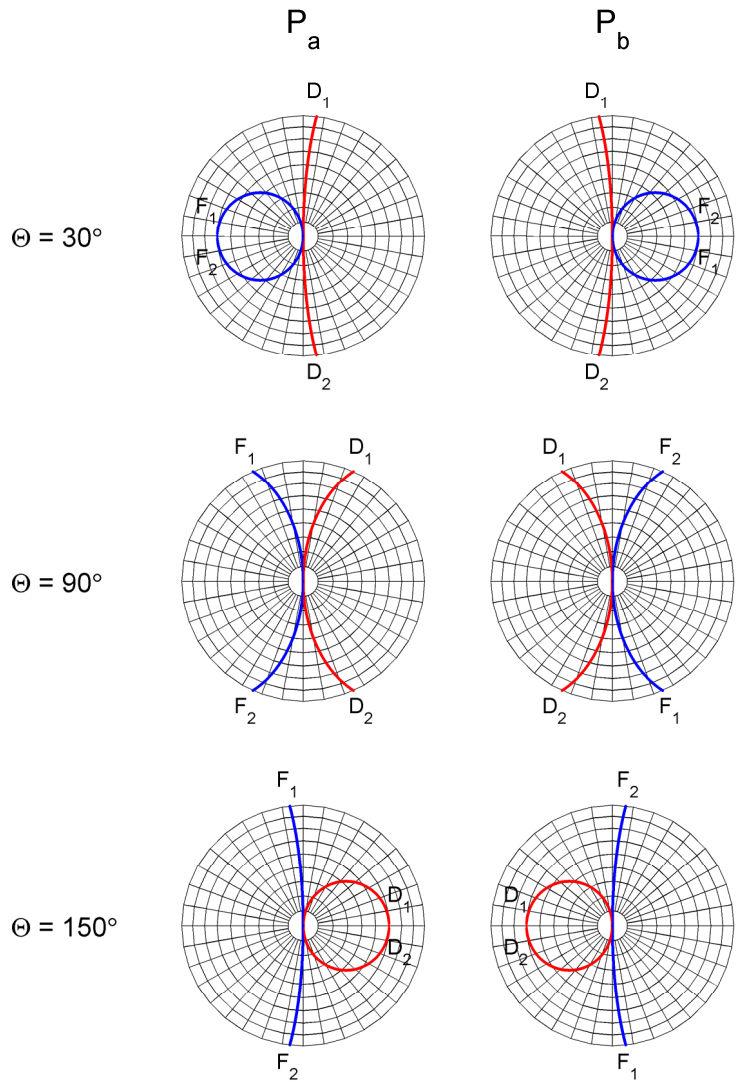

**Figure 3** Common intersection arcs of two Ewald sphere-projected scattering patterns ($P_a$, $P_b$) extending to 45° scattering angle for several $\Theta$ hinge angles. It is assumed that the $\Phi$ and $\Psi$ azimuthal rotations already applied to bring the tangent common line to the same (vertical) direction. The red curves show the two branches of the "direct" intersection arcs ($D_1$, $D_2$), and the blue ones show the "Friedel" intersection arcs ($F_1$, $F_2$). Note the flipped branches of the Friedel arcs.

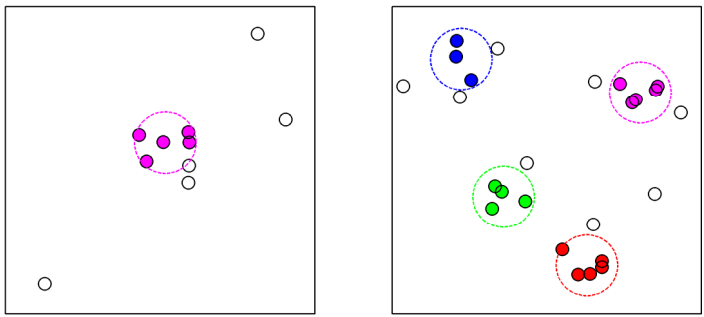

**Figure 4** Simple illustration of selection and averaging procedure in 2D. Points on the plane represent orientations; their distance corresponds to the misorientation angle. A group quality threshold, i.e. a radius is given as a parameter, and the neighbours within this radius are determined for each elements. The most populated neighborhood is selected and averaged. Left panel: Only one group is determined (solid circles), the remaining elements are treated as outliers (empty circles). Right panel: The procedure is applied repeatedly to obtain a predefined number of groups (various colors), in between excluding the already clustered elements.

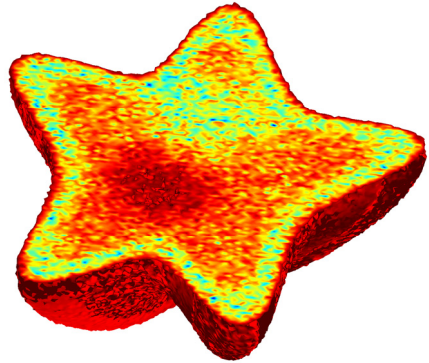

**Figure 5** Absolute value of the 3D complex scattering density of the virus model shown on a combined isosurface and cut-slice plot. The object has a pseudo symmetric shape, internal structure on various length scales with small contrast. An imaginary part was added to make Friedel symmetry invalid.

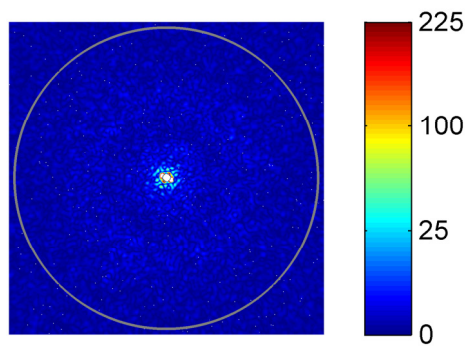

**Figure 6** One of the 100 calculated scattering patterns of the RuBisCo protein. The square root intensity scale represents photon counts in the Shannon-Nyquist sampling pixel size, having $(\lambda/2D)^2$ solid angle. The average count in the indicated resolution shell of 2.4 Å is about 0.24. The mask of the missing and saturated pixels visible in the center as white areas is handled by the orientation algorithm.

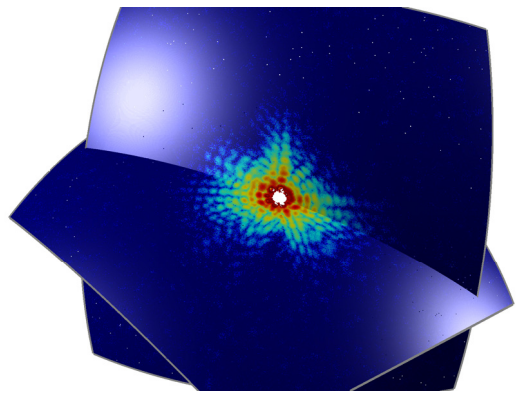

**Figure 7** Zoomed intersection of 2 oriented scattering patterns of the virus model showing the perfect correlation of intensities along the common arc. The hole in the center is due to the saturated and missing pixels that were made transparent.

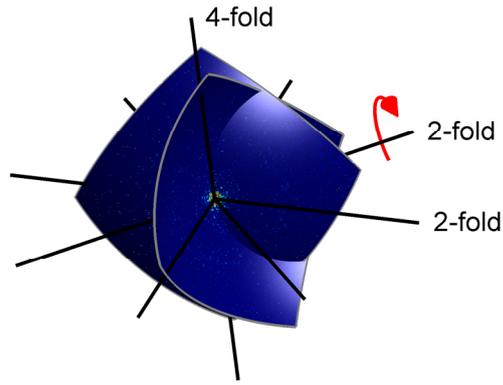

**Figure 8** 2 possible settings of a pattern and the 4- and 2-fold axes of symmetry operations as found by the Common Arc orientation method. The 2 settings are found to be related by a 180º rotation (indicated by the red arrow) that defines one of the 2-fold axes. All equivalent settings of all patterns define the symmetry operations with high precision. These form the RuBisCos's complete 422 point group. Plotting of all 8 equivalent and also the 8 Friedel inverted settings of the oriented pattern would make the figure too busy.

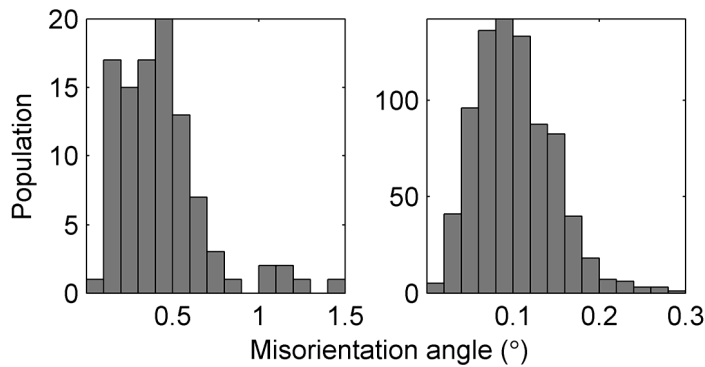

**Figure 9** Distribution of the misorientation angles of the oriented patterns for the virus model object (left panel) and the RuBisCo protein assembly (right panel). The values verify the orientations determined by the Common Arc method against the original ones kept from the data preparation. Low average and maximum misorientation angles prove perfect orientation.

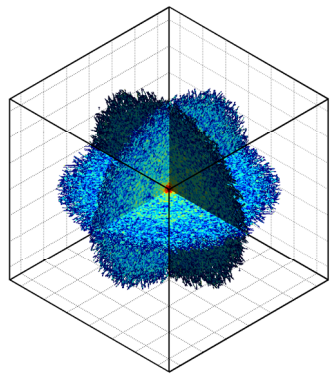

**Figure 10** 3 main slices of the 3D intensity distribution in the reciprocal space as constructed from the original 2D scattering patterns of RuBisCo oriented by the Common Arc method. The fuzzy surface of the roughly spherical volume is due to the voxels that do not have contribution from any of the oriented patterns. Voxels further in are defined by more and more patterns with higher reliability. This data has a correlation of 0.98 with the original intensity distribution in the reciprocal space.